# Detection of Structural Regimes and Analyzing the Impact of Crude Oil Market on Canadian Stock Market: Markov Regime-Switching Approach

Mohammadreza Mahmoudi[a] , Hana Ghaneei[b]

[a] *Department of Economics, Northern Illinois University, Dekalb, USA.*

Email: mmahmoudi@niu.edu

[b] *Department of Industrial and Systems Engineering, Northern Illinois University, Dekalb, USA.*

Email: z1939548@students.niu.edu

**Correspondence**

Mohammadreza Mahmoudi, *Northern Illinois University, Dekalb, USA*.

Email: mmahmoudi@niu.edu

**Abstract**

This study aims to analyze the impact of the crude oil market on the Toronto Stock Exchange Index (TSX)c based on monthly data from 1970 to 2021 using Markov-switching vector autoregressive (MSI-VAR) model. The results indicate that TSX return contains two regimes, including: positive return (regime 1), when growth rate of stock index is positive; and negative return (regime 2), when growth rate of stock index is negative. Moreover, regime 1 is more volatile than regime 2. The findings also show the crude oil market has negative effect on the stock market in regime 1, while it has positive effect on the stock market in regime 2. In addition, we can see this effect in regime 1 more significantly in comparison to regime 2. Furthermore, two period lag of oil price decreases stock return in regime 1, while it increases stock return in regime 2.

1. **Introduction**

The impact of the financial crisis in 2007 on the global economy strongly proves the pivotal role of the stock market on the economy. The stock market through different financial instruments could provide diverse funding for companies and motive economic growth. These positive effects of the stock market in developed countries like Canada are more obvious. Canada is the world's second-largest country by total area, and its advanced economy is the ninth-largest in the world. Moreover, the country in question has ranked sixteenth among the countries with the highest nominal per-capita income and come out sixteenth in the Human Development Index around the world.[1] As summarized in Table 1, market capitalization of listed domestic companies for Canada was 2.64 trillion dollars in 2020. The percent of market capitalization of listed domestic companies in the economy (160.7%) and stock traded in percent of Gross Domestic Product (GDP) (61.5%) in 2020 shows the important role of the stock market in Canadian economy. In addition, the Toronto Stock Exchange Index (TSX) as the Canada's main stock exchange with 3.3 trillion total market capitalization of 1,640 listed companies up until August 2021 is the tenth-largest stock exchange in the world by market capitalization.[2]

---

[1] See the latest Human Development Index ranking which is available at http://hdr.undp.org/en/content/latest-human-development-index-ranking
[2] Visit the Toronto Stock Exchange website at https://www.tsx.com/listings/listing-with-us

Table 1. Economic data of Canada in 2020

| | | Quantity/Percent | Rank in the World |
|---|---|---|---|
| **Macroeconomics** | GDP (constant 2010 trillion USD) | 1.847 | 9 |
| | GDP growth (annual %) | -5.4 | |
| | Population | 38,005,238 | 37 |
| | Population growth (annual %) | 1.1 | |
| | GDP per capita (constant 2010 USD) | 48,617.09 | 20 |
| | GDP per capita growth (annual %) | -6.6 | |
| **Stock Market** | Market Capitalization of listed domestic companies (trillion USD) | 2.64 | 6 |
| | Market Capitalization of listed domestic companies (% of GDP) | 160.7 | |
| | Total value of stocks traded (trillion USD) | 1.01 | |
| | Total value of stocks traded (% of GDP) | 61.5 | |
| | Listed domestic companies | 3,922 | |
| **Oil Market** | Crude Oil Export (billion USD) | 47.6 | 6 |
| | Crude Oil Export (% of total value of Canada's exports) | 12.2 | |
| | Crude Oil Export (% of total value of exported crude oil in the World) | 5.9 | |

*Note: This table summarized the main economic data of macroeconomics, stock market, and oil market of Canada in 2020 based on the Word Bank database*

The stock price index reveals the general trend/tendency of the stock market fluctuations. Indeed, the degree to which capital market is likely to succeed or fail is understood by way of the trend/tendency. Accordingly, in order for the portfolio managers and investors, who are dealing with stock trading and other financial assets in the market, to maintain and increase their value of portfolio, they need several factors influencing the stock price under discrepant economic conditions.

Moreover, the returns of the stock exchange are under the influence of various factors. Oil price stands out among these affecting factors. Oil and its products are being used as the major source of energy in manufacturing processes across the world. Therefore, fluctuations in oil price can have a direct impact on the cost of production and company profitability. For example, while Gurrib (2021) and Prabheesh, Padhan, and Garg (2020) recently found that shocks to the number of crude oil price volatility impacted stocks during early COVID-19, Gurrib (2018), Gurrib (2019) and Lescaroux and Mignon (2008) find crude oil prices to be poor predictors in other markets. This is in line with Kinateder, Campbell, and Choudhury

(2021) who compared correlations across different asset classes and found deteriorating relationships. Bakas and Triantafyllou (2020) found that uncertainty related to COVID-19 had a negative impact on crude oil.

Oil is the most vital source of profit for oil exporting countries like Canada, Iran, Qatar, and the United Arab Emirates, which is why, its price and fluctuations can influence the natural sector and the capital market Mahmoudi (2021). Based on Table 1, more than 12 percent of total value of Canada's exports was crude oil in 2020, so that the biggest export products by value of Canada were crude oil. In addition, Canada exported 47.6 billion dollars crude oil in 2020, which was 5.9 percent of total value of exported crude oil in the World. Canada is the sixth-largest producer and exporter of oil in the world. Even though some studies recommend that Canada needs to invest in new sources of energy like wave energy, solar energy, and wind energy in order to reduce the Canadian economy's dependence on oil revenue, therefore, in this situation oil market fluctuations have less negative effect on the economy. (Ghaneei and Mahmoudi 2021)

All in all, the oil market plays a pivotal role in the economics of Canada. In addition, nowadays stock markets have inevitable role in growth economics of developed country such as Canada. Because of these issues we want to analyze the effect of oil market on stock market in Canada. In this paper, we use a Markov-switching var model to examine this relationship. First, we introduce some important papers, which use different Markov-switching models, to analyze the relationship between oil price and the stock market. Then, we use data and run the MSI-VAR model to find the best fitted model to show the relation between TSX and West Texas Intermediate Crude Oil (WTI). Finally, we discuss main results and provide conclusion.

2. Literature Review

Among the existing models employed in computational economics and econometric time-series analysis regime switching models have proved the most preferable. These models, for the most part, consist in a two-regime pattern with a state process which determines, in each period, the occurrence of one of the regimes. This state process, which is bivalued, frequently uses Markov Chain as an example to be modeled upon. Hamilton (1988) opted for this category of Markov-switching in the mean to carry out the Autoregressive model. In Hamilton (1989) , a two-state Markov-switching model in which the mean growth rate of Gross national product (GNP) is subject to regime switching was specified, where the errors followed a regime-invariant AR(4) process. The Hamilton method was taken up by Kim (1994) and became the subject of a more detailed analysis. Markov-switching was classified more generally, on the part of various authors, in the category of such models as regression models and volatility models. Filardo (1994) used an MSAR(4) switching mean model to simulate the log growth rate of industrial production (DLOGIP). He used The log growth rate, which was derived from the Composite Index of Eleven Leading Indicators (DLOGIDX), was made use of by him to serve as a business-cycle predictor. Krolzig (2000) studied the common business cycle of six Organization for Economic Co-operation and Development (OECD) countries across four continents by using Markov-switching Vector Autoregressions (MSVAR). Moreover, various statistical characteristics of the Markov-switching model have been analyzed by Garcia (1998), Timmermann (2000), Cho and White (2007) .

On the nature of the link formed between the oil market and the stock market in various countries. Park and Ratti (2008) conducted a study in the USA and thirteen other developed countries in which they investigated the degree to which fluctuations in oil market price influenced such pivotal financial factors as stock prices. The result revealed that there, statistically, existed a significant correlation between the fluctuation in oil price and stock index. Nevertheless, to determine the type of the correlation much hinges upon the fact whether the countries in question are oil exporters or importers. In a study Fayyad and Daly (2011) examined, comparatively, the impact the oil prices shocks exerted on the stock profits in the Gulf Cooperation Council (GCC)[3] countries, the United States, and the United Kingdom. They found that the rise in oil price during the global crisis increased the predictive power of oil prices on the stock market.

---

[3] Gulf Cooperation Council (GCC) is a political and economic union that consists of the United Arab Emirates, Saudi Arabia, Kuwait, Bahrain, Qatar, and Oman. For more information visit https://www.worldbank.org/en/country/gcc/overview#1

Moreover, Qatar, the United Arab Emirates and the UK are among the countries whose stock markets showed more sensitivity to the shocks. Hussin et al. (2013) conducted an in-depth analysis of the link between the strategic goods, such as gold price and oil price, and the stock market in the context of Malaysia by making use of such models as Vector Auto Regression (VAR), Correlation Analysis, Granger, Causality Test, Impulse Response Function (IRF) and Variance Decomposition (VDC). The result demonstrated that, in the long run, stock profits do not form an integrated relationship with the strategic goods. Furthermore, the Granger causality model shed light on the fact that in Malaysia, of all the strategic goods, only oil price variables will have impact on stock returns in the short run. Balcilar, Gupta, and Miller (2015) conducted a comprehensive analysis in the USA spanning more than a half a century (monthly data from late 1950s to early 2010s) to study the non-linear relationship between oil price and stock market price using a MS-VEC model. The result revealed that there had been a model of two regimes in effect that splits the sample into two regimes of high- and low-quality pivoting around the variance-covariance matrix of oil and stock prices. Moreover, stock prices and oil price shocks have negative correlation in the high-volatility regime, while there is no relationship between them in the low-volatility regime.

3. **Methodology**

The general MS-VAR model of order p with m regime for a time series variable $y_t$ is shown as follows:

$$y_t = \begin{cases} v_1 + a_{11}y_{t-1} + \ldots + a_{p1}y_{t-p} + \Sigma_1^{1/2}u_t, & \text{if } s_t = 1 \\ v_m + a_{1m}y_{t-1} + \ldots + a_{pm}y_{t-p} + \Sigma_m^{1/2}u_t, & \text{if } s_t = m \end{cases}$$

$$u_t|s_t \sim NID(0, I_k) \qquad (1)$$

We could write the above equation as follows:

$$y_t = v_{s_t} + \sum_{i=1}^{p} a_{i,s_t} y_{t-i} + \Sigma_{s_t}^{1/2} u_t \qquad (2)$$

Where $y_t = \begin{pmatrix} y_1 \\ \vdots \\ y_p \end{pmatrix}$, $a_i = \begin{pmatrix} a_{11} & \cdots & a_{1p} \\ \vdots & \ddots & \vdots \\ a_{p1} & \cdots & a_{pp} \end{pmatrix}$, $y_{t-i} = \begin{pmatrix} y_{t-1} \\ \vdots \\ y_{t-p} \end{pmatrix}$, $v_{S_t} = \begin{pmatrix} v_{1,S_t} \\ \vdots \\ v_{1.S_t} \end{pmatrix}$

$a_i$ indicates the matrix of autoregressive coefficients. $v$ is intercept term, $\Sigma^{1/2}$ is the standard deviations which depends on $s_t$, the regime at time t.

It is convenient to assume that in equations 1 and 2, error term, $u_t$, normally distributed, hence the conditional probability density function of $y_t$ is as follows:

$$p(y_t|s_t = t_m, Y_{t-1}) = \ln\left(2\pi^{-\frac{1}{2}}\right) \ln\left|\Sigma^{-\frac{1}{2}}\right| \exp\{(y_t - \overline{y_{mt}})'\Sigma_m^{-1}(y_t - \overline{y_{mt}})\} \quad (3)$$

where $\overline{y_{mt}} = E[y_t|s_t, Y_{t-1}]$, and $Y_{t-1} = (y'_{t-1}, y'_{t-2}, \ldots, y'_1, y'_0, \ldots, y'_{1-p})'$

Therefore, the conditional density function of $y_t$ is normal, $y_t|s_t = t_m, Y_{t-1} \sim NID(\overline{y_{mt}}, \Sigma_m)$

$$\overline{y_t} = \begin{bmatrix} \overline{y_{1t}} \\ \vdots \\ \overline{y_{mt}} \end{bmatrix} = \begin{bmatrix} v_1 + \sum_{i=1}^{p} a_{1,i} y_{t-i} \\ \vdots \\ v_m + \sum_{i=1}^{p} a_{1,m} y_{t-i} \end{bmatrix} \quad (4)$$

In the above equations, we assumed the state variables, $s_t$, which specify the switching behavior of time series variable, $y_t$, follows an irreducible ergodic two-state Markov process. This assumption indicates that a current regime $s_t$ depends on the regime one period ago, $s_{t-1}$. Hence, the transition probability between states is as follows:

$$Pr(S_t = j|S_{t-1} = i, S_{t-2} = k, \ldots) = Pr(S_t = j|S_{t-1} = i) = p_{ij} \quad (5)$$

$p_{ij}$ denotes the transition probability from state $i$ to state $j$.

Generally, the transition probability is identified by a ($n \times n$) matrix as follows:

$$P = \begin{bmatrix} p_{11} & p_{12} & \cdots & p_{1m} \\ p_{21} & p_{22} & \cdots & p_{2m} \\ \vdots & \vdots & \ddots & \vdots \\ p_{m1} & p_{m2} & \cdots & p_{mm} \end{bmatrix} \quad (6)$$

$$\sum_{j=1}^{m} p_{ij} = 1, i = 1, 2, \ldots, m, \text{ and } 0 \leq P_{ij} \leq 1$$

There are many Markov-switching vector auto regression models that we could use to analyze the effect of the oil market on the stock market. In **Error! Reference source not found.** these models are summarized. All in all, we have two main Markov-switching var models, Markov-switching mean and Markov-switching intercept. It should be noted that, MSM and MSI models are different. The rapid leap of the observed time series vector onto a new level is brought about by the dynamic response to the regime shift in the mean $\mu(s_t)$. On the other hand, a regime shift in the intercept term $\nu(s_t)$ cause a dynamic response which is quite the same as a comparable shock in the white noise series $u_t$.

Table 2. Different Markov-switching VAR models

|  |  | MSM | | MSI | |
| --- | --- | --- | --- | --- | --- |
|  |  | $\mu$ invariant | $\mu$ varying | $\nu$ invariant | $\nu$ varying |
| $A_j$ invariant | $\Sigma$ invariant | MSM-VAR | Linear MVAR | MSI-VAR | Linear VAR |
|  | $\Sigma$ varying | MSMH-VAR | MSH-MVAR | MSIH-VAR | MHA-VAR |
| $A_j$ varying | $\Sigma$ invariant | MSMA-VAR | MSA-MVAR | MSIA-VAR | MSA-VAR |
|  | $\Sigma$ varying | MSMAH-VAR | MSAH-MVAR | MSIAH-VAR | MSAH-VAR |

*Note: This table contains all Markov-switching Vector Autoregressive Models. M indicates Markov-switching mean, I represents Markov-switching intercept, A shows Markov switching autoregressive, and H determines Markov-switching heteroskedasticity.[ Krolzig (1998)]*

## 4. DATA

In this study we use monthly data for TSX from January 1970 to May 2021. Moreover, so as to analyze the effect of oil prices on the stock market we use monthly data of closing spot prices for West Texas Intermediate (WTI) crude oil from January 1970 to May 2021. WTI serves as one of the main pricing references for light crude oil in the United States as well as it is a benchmark crude oil for the North American market.[4] The data for TSX is obtained from Statistics Canada[5] and the data for WTI is sourced from the US Energy Information Administration (EIA).[6] Since several important oil price shock like 1973 and 1979 oil crisis, 1990 oil price shock, 2000s energy crisis, 2007-2008 financial crisis, 2020 Russia–

---

[4] Based on several analysis the best proxy for Western Canada Select (WCS) is WTI. Since, there is limited dataset for WCS price, we use WTI price instead. See the Alberta Economic Dashboard at https://economicdashboard.alberta.ca/oilprice
[5] Visit https://www150.statcan.gc.ca/n1/en/type/data?MM=1#tables to get the data for TSX
[6] Visit https://www.eia.gov/dnav/pet/hist/LeafHandler.ashx?n=PET&s=RWTC&f=M to get the data for WTI.

Saudi Arabia oil price war, and Covid-19 pandemic happened during January 1970 to May 2021, this range of data helps us to get a comprehensive understanding regarding the impact of oil price shocks on the Canadian stock market CLAYTON (2015)

Some of the main statistics of data are described in Table 3. The mean and standard deviation of TSX are 6807.55 and 5234.50 respectively. The maximum index was 19730.99, which was reported in May 2021, and the minimum was 810.78, which was reported in June 1970. However, the WTI has experienced less volatility. The mean and standard deviation of WTI are 36.32 and 27.59 respectively. The minimum quantity was 3.31 which was reported in July 1970, and the maximum quantity was 133.88 which was reported in June 2008.

*Table 3. Summary Statistics of Natural Logarithm of levels and returns for TSX and WTI*

|  | TSX | WTI |
|---|---|---|
| **Panel A: Natural Log Levels** | | |
| Mean | 6807.55 | 36.32 |
| Median | 4554.85 | 27.76 |
| Maximum | 19730.99 | 133.88 |
| Minimum | 810.78 | 3.31 |
| Standard Deviation | 5234.50 | 27.59 |
| Skewness | 0.54 | 1.14 |
| Kurtosis | -1.14 | 0.53 |
| Jarque-Bera | 47.33 | 34.92 |
| Observation | 617 | 617 |
| | | |
| **Panel B: Natural Log Returns** | | |
| Mean | 0.0048 | 0.0048 |
| Median | 0.0086 | 0.00 |
| Maximum | 0.16 | 0.85 |
| Minimum | -0.26 | -0.57 |
| Standard Deviation | 0.046 | 0.092 |
| Skewness | -1.05 | 0.65 |
| Kurtosis | 7.30 | 21.62 |
| Jarque-Bera | 588.58 | 8943.39 |
| Observation | 616 | 616 |

*Note: Panel A summarized the main descriptive statistics of natural logarithms of levels for TSX and WTI. Panel B summarized the main descriptive statistics of returns for TSX and WTI.*

It should be noted that our analysis is based on the calculated return ($return = ln(\frac{P_t}{P_{t-1}})$) for all variables. In Plot 1, we show the graph for natural logarithms of TSX and WTI on the left side and we depict the returns of the TSX and WTI on the right side.

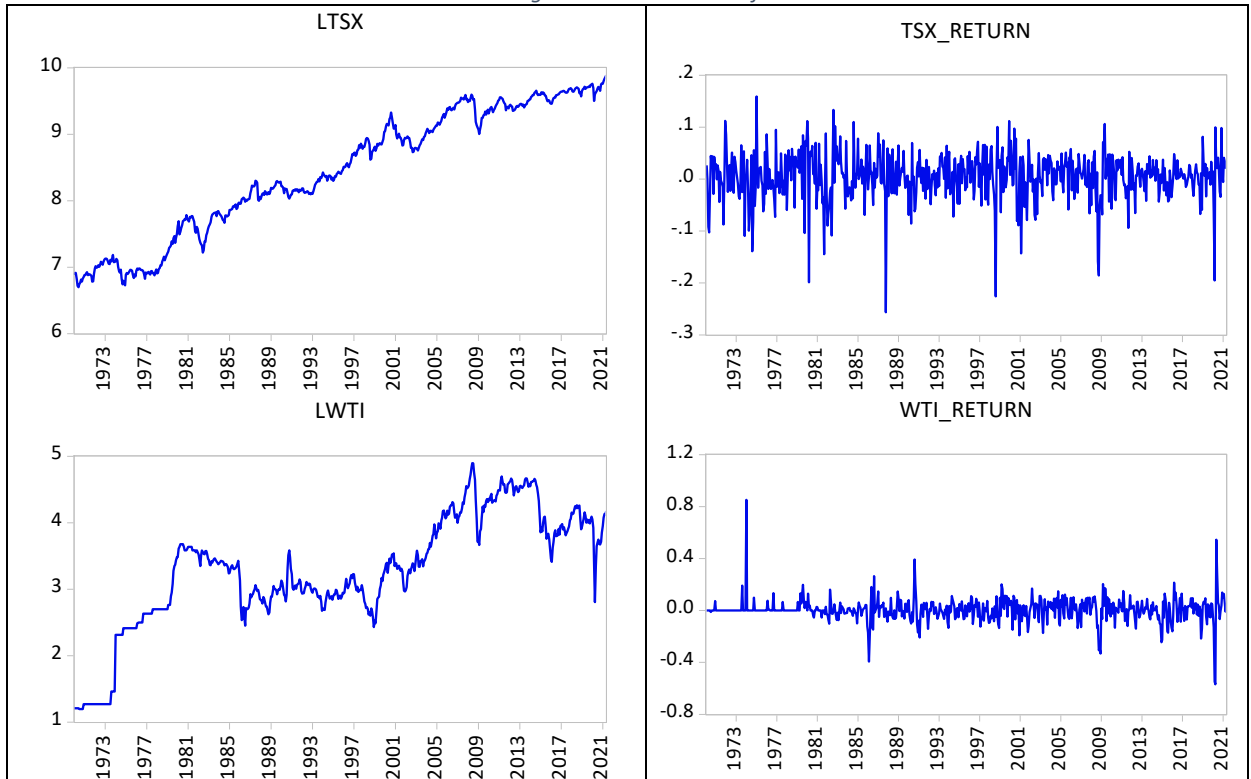

*Plot 1. Natural logarithms and returns of TSX and WTI*

*Note: The left panels show the plot of natural logarithms of TSX and WTI as well as the right panels depict the returns of TSX and WTI.*

## 5. Empirical Results and Discussion

we tested stationarity using Augmented Dickey Fuller (ADF), Phillips–Perron, and Kwiatkowski-Philips-Schmidt-Shin (KPSS). Based on Table 4, all the time series data are non-stationary in the natural log level and stationery in the return level.

*Table 4. Unit Root Test for Natural Log levels and Returns of TSX and WTI*

|                 | LTSX    | TSX Return | LWTI    | WTI Return |
|-----------------|---------|------------|---------|------------|
| **ADF**         | -3.303  | -22.323    | -3.132  | -19.319    |
|                 | (0.067) | (0.000)    | (0.099) | (0.000)    |
| **Phillips–Perron** | -3.281  | -22.327    | -2.769  | -18.792    |
|                 | (0.071) | (0.000)    | (0.209) | (0.000)    |
| **KPSS**        | 0.373   | 0.025      | 0.190   | 0.048      |

*Note: This table summarized ADF, Phillips–Perron, and KPSS unit root tests in order to test stationarity of natural logarithms and returns of TSX and WTI. The results shows that variables are non-stationary in the natural log level and stationery in the return level. The value in parentheses represent p-value.*

Next, we wanted to recognize a long-run relationship between our return variables. In order to test cointegration in our data we should first find optimal lag. We ran the VAR model to find optimal lag. Vector autoregression (VAR) is a stochastic process model used to comprehend the linear interconnections among various time series. The univariate autoregressive model, known as AR model, is generalized via VAR models by means of introducing multiple evolving variables.

For example VAR(p) Model is as follows:

$$Y_t = a + A_1 Y_{t-1} + A_2 Y_{t-2} + \ldots + A_p Y_{t-p} + \varepsilon_t$$

As it is clear from Table 5, based on the AIC criteria the optimal lag length is 2.

*Table 5. VAR lag order selection criteria*

| Lag | LogL | LR | FPE | AIC | SC | HQ |
|---|---|---|---|---|---|---|
| 0 | -1267.623 | NA | 0.223272 | 4.176391 | 4.190899 | 4.182035 |
| 1 | 1611.688 | 5730.208 | 1.74e-05 | -5.281869 | -5.238348 | -5.264937 |
| 2 | 1640.615 | 57.37769* | 1.61e-05* | -5.363865* | -5.291330* | -5.335645* |
| 3 | 1643.027 | 4.767829 | 1.61e-05 | -5.358641 | -5.257090 | -5.319133 |
| 4 | 1646.411 | 6.668267 | 1.62e-05 | -5.356615 | -5.226051 | -5.305819 |
| 5 | 1648.082 | 3.282128 | 1.63e-05 | -5.348955 | -5.189376 | -5.286871 |
| 6 | 1649.923 | 3.603331 | 1.64e-05 | -5.341853 | -5.153260 | -5.268481 |
| 7 | 1652.741 | 5.496625 | 1.65e-05 | -5.337964 | -5.120357 | -5.253304 |
| 8 | 1656.867 | 8.021523 | 1.65e-05 | -5.338379 | -5.091757 | -5.242431 |

*Note: This table summarized several criteria to determine the optimal lag length of VAR model. Based on different criteria, the optimal lag for our data is 2.*

After finding the optimal lag, we can run the cointegration test. The Johansen test is a test for evaluating cointegration and it allows to test more than one cointegrating relationship. Johansen test estimates the rank (r) of a given matrix of time series with a confidence level. Based on Table 6, our data set has two time series, hence, Johansen tests the null hypothesis of $r = 0$ in which there is no cointegration at all, and the null hypothesis of $r \leq 1$ in which there is at most one cointegration relation. For $r = 0$, all of the test values are less than the critical values, so we could not reject the null hypothesis and there is no cointegration relation at all. For $r \leq 1$, since none of the test values are greater than the critical values, the null hypothesis is rejected. Therefore, there is no cointegration relation between our variables.

*Table 6. Johansen cointegration test*

|  | Test Statistics | Critical Values | | |
|---|---|---|---|---|
|  |  | 10 percent | 5 percent | 1 percent |
| **Panel A: Trace Test** | | | | |
| $r = 0$ | 5.95 | 7.52 | 9.24 | 12.97 |
| $r \leq 1$ | 16.04 | 17.85 | 19.96 | 24.60 |
| **Panel B: Maximum Eigenvalue** | | | | |
| $r = 0$ | 5.95 | 7.52 | 9.24 | 12.97 |
| $r \leq 1$ | 10.08 | 13.75 | 15.67 | 20.20 |

*Note: This table summarized the Trace and Maximum Eigenvalue tests in order to evaluating cointegration relationships in our data. The results of different tests show there is no cointegration relations between TSX and WTI.*

In absence of cointegration, the dependent variables diverge from any possible combination of the regressors. Therefore, running a VAR in levels is not reasonable. If certain or every single variable in a regression rate is I (1) therefore the regular statistical results might or might not be maintained. Spurious regression is a seminal example of when the regular statistical results are not maintained, while every single regressor is I (1) and not cointegrated as well. To avoid spurious regression and to seek nonlinearities in oil prices and the stock index relationship, we use Markov-switching VAR model. Based on the AIC criteria, the best model which fits the data is MSI(2)-VAR(2) model.

For analyzing our data, we ran MSI(2)-VAR(2) model to identify TSX index's structure. As it is shown in Table 7, TSX index return has two regimes including positive return (regime 1), when the growth rate of stock index is positive, and negative return (regime 2), when the growth rate of stock index is negative.

*Table 7. MSI(2)-VAR(2) Model of TSX Index return*

| | | $DLN(TSX)_t$ |
|---|---|---|
| **Intercepts** | **Const (Regime 1)** | 0.01 (0.00) |
| | **Const (Regime 2)** | -0.015 (0.00) |
| **Autoregressive Coefficients** | **AR(1)** | 0.022 (0.58) |
| | **AR(2)** | -0.024 (0.55) |

*Note: This table summarized the results of TSX index's structure using MSI(2)-VAR(2) Model. $DLN(TSX)_t$ indicates natural logarithms of return of TSX at time t. The results show TSX index return has two regimes including positive return (regime 1) and negative return (regime 2). The values in parenthesis represent p-value.*

Then, we computed the transition probabilities of MSI(2)-VAR(2) Model for TSX index return. As it is shown in Table 8, $P_{11}$, which represents the probability of staying in regime 1 given that you are in regime 1, is 96%, as well as $P_{21}$, which indicates the coefficients for the transition from regime 2 into regime 1, is 17%.

*Table 8.Transition Probabilities of MSI (2)-VAR (2) Model for TSX Index*

|  | Regime 1 | Regime 2 |
|---|---|---|
| **Regime 1** | 0.959781 | 0.040219 |
| **Regime 2** | 0.167467 | 0.832533 |

*Note: This table summarized the transition probabilities of TSX index using MSI (2)-VAR (2). For example, $P_{11}$, the probability of staying in regime 1 given that you are in regime 1, is 96%, and $P_{21}$, the coefficients for the transition from regime 2 into regime 1, is 17%.*

The filtered probabilities of being in two regimes for TSX index is depicted in the Plot 2.

*Plot 2.Markov-switching Filtered Regime Probabilities of TSX index return*

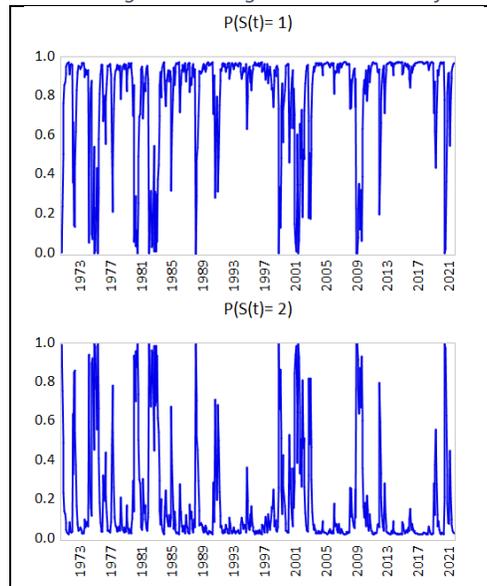

*Note: This plot shows Markov-switching filtered probabilities of regime 1 and 2 for TSX index return. regime 1 includes positive return and Regime 2 contains negative returns.*

Based on the AIC criteria, MSI(2)-VAR(2) Model is the best fitted model for examining the effect of the crude oil market on the stock market. According to Table 9 information, in regime 1 the expected return of TSX index increases monthly by 0.18 and the expected return of WTI price rises monthly by 0.57, while in regime 2 the expected return of TSX index decreases monthly by 0.02 and the expected return of WTI price rises by 0.003. In addition, the standard deviation in regime 1 is higher than regime 2, which indicates the

TSX bull market is more volatile than the TSX bear market. Moreover, these results imply that there exists a negative correlation between oil price and stock return in regime 1, however, this relationship is positive in regime 2. In fact, in TSX bull market when oil price increases the stock return decrease by 0.177, while in bear market when oil price rises the stock return increases by 0.017. Furthermore, two period lag of oil price decreases the stock price in regime 1, while it increases the stock price in regime 2.

Table 9. The effect of oil price on stock index with MSI (2)-VAR (2) Model (Based on AIC criteria)

|  |  | $DLN(TSX)_t$ | $DLN(WTI)_t$ |
|---|---|---|---|
|  | **Intercepts** |  |  |
|  | **Const (Regime 1)** | 0.1893 (0.111) | 0.5773 (0.2543) |
|  | **Const (Regime 2)** | -0.0202 (0.0491) | 0.003507 (0.3049) |
|  | **Standard Deviation** |  |  |
|  | **Sigma (Regime 1)** | 0.005 | 0.001 |
|  | **Sigma (Regime 2)** | 0.001 | 0.0001 |
|  | **Coefficients** |  |  |
| **Regime 1** | $DLN(TSX)_{t-1}$ | -0.0216 (0.051) | 0.4120 (0.1226) |
|  | $DLN(TSX)_{t-2}$ | 0.0337 (0.053) | -0.1750 (0.119) |
|  | $DLN(WTI)_{t-1}$ | -0.1772 (0.121) | -0.3271 (0.269) |
|  | $DLN(WTI)_{t-2}$ | -0.012 (0.007) | -0.002 (0.015) |
| **Regime 2** | $DLN(TSX)_{t-1}$ | 0.021 (0.024) | 0.088 (0.041) |
|  | $DLN(TSX)_{t-2}$ | -0.025 (0.022) | -0.015 (0.035) |
|  | $DLN(WTI)_{t-1}$ | 0.017 (0.046) | 0.026 (0.067) |
|  | $DLN(WTI)_{t-2}$ | 0.009 (0.001) | 0.006 (0.002) |

*Note: This table summarized the effect of oil market on Canadian stock market using MSI(2)-VAR(2). $DLN(TSX)_t$ and $DLN(WTI)_t$ indicate natural logarithms of return of TSX and WTI at time t. The results show that there exists a negative correlation between oil price and stock return in regime 1, however, this relationship is positive in regime 2. It means, increasing oil price decreases stock return in bull market, while increasing oil price increases stock return in bear market. Moreover, two period lag of oil price decreases the stock price in regime 1, while it increases the stock price in regime 2. In addition, regime 1 is more volatile than regime 2. The values in parenthesis represent standard error.*

Furthermore, the transition probabilities of MSI(2)-VAR(2) Model for the effect of oil market on Canadian Stock Index is shown in Table 10. Based on this information, when we consider the effect of oil prices, $P_{11}$, which represents the probability of staying in regime 1 given that you are in regime 1, is 96%, as well as $P_{21}$, which indicates the coefficients for the transition from regime 1 into regime 2, is 17%.

*Table 10.Transition Probabilities of MSI(2)-VAR(2) Model for the effect of oil market on Canada Stock Index*

|  | Regime 1 | Regime 2 |
|---|---|---|
| **Regime 1** | 0.961540 | 0.038460 |
| **Regime 2** | 0.178614 | 0.821386 |

*Note: This table summarized the transition probabilities of TSX index using MSI (2)-VAR (2). For example, $P_{11}$, which represents the probability of staying in regime 1 given that you are in regime 1, is 96%, as well as $P_{21}$, which indicates the coefficients for the transition from regime 1 into regime 2, is 17%.*

## 6. Conclusion

Based on the unit root test, all variables are non-stationary in the level and stationary in the first difference. Also, the cointegration test indicates that there is no long-run relation between the crude oil market and the stock index. Hence, we cannot run a VAR model because of the spurious regression. To avoid spurious regression and in order to examine the nonlinear relationship between oil prices and the stock index we used a Markov regime switching model. Based on the MSI(2)-VAR(2) model there are two regimes for the Canadian stock index including positive return (regime 1), when growth rate of stock index is positive, and negative return (regime 2), when growth rate of stock index is negative. Moreover, this model shows the crude oil market has positive effect on the stock market in both regimes, however, the oil price affects the stock market in regime 1 is more notably than regime 2. Furthermore, a two-period lag of oil price increases stock price in regime 1, while it decreases stock price in regime 2.